\documentclass[letter]{aa} 

\usepackage{graphicx}
\usepackage{txfonts}
\usepackage[pdftex, colorlinks=true, linkcolor=blue, citecolor=blue, urlcolor=blue]{hyperref}
\usepackage{nameref} 
\usepackage{ulem}
\newcommand{\mic}{$\mu$m}
\newcommand\rev[1]{#1}
\begin{document}

   \title{The role of absorption and scattering in shaping ice bands. Spatially-resolved spectroscopy of protoplanetary disks}


   \author{L. Martinien\inst{1}
          \and
          G. Duch\^ene\inst{1,2}
          \and
          F. M\'enard\inst{1}
          \and 
          R. Tazaki\inst{3}
          \and 
          K.R. Stapelfeldt\inst{4,}\thanks{On sabbatical leave at IPAG}
          }

   \institute{Univ. Grenoble Alpes, CNRS, IPAG, 38000 Grenoble, France \\
    \email{laurine.martinien@univ-grenoble-alpes.fr}
    \and Astronomy Department, University of California Berkeley, Berkeley CA 94720-3411, USA
    \and Department of Earth Science and Astronomy, The University of Tokyo, Tokyo 153-8902, Japan
    \and Jet Propulsion Laboratory, California Institute of Technology, 4800 Oak Grove Drive, Pasadena, CA 91109, USA
             }

   \date{}

 
  \abstract
   {} 
   {The James Webb Space Telescope now enables the spectral study of ices with unprecedented sensitivity and angular resolution. Water ice plays a crucial role in the growth of grains and in planetary formation but its spatial distribution in protoplanetary disks is poorly constrained. To help the interpretation of future observations, we study here for the first time how the water ice band depends on the observer's perspective and the location where spectra are measured within protoplanetary disks.}
   {Based on a standard protoplanetary disk model around a T Tauri star, we used the radiative transfer code MCFOST to extract water-ice spectra and to measure the depth and central wavelength of the water-ice band at different locations in the disk.}
   {Even in the context of a spatially homogeneous ice mixture, the observed properties of water-ice bands depend on the inclination of the system as well as on the location in the disk from which the spectra are extracted. In particular, the wavelength of the band minimum can change by up to 0.17\,\mic\,, comparable to the difference expected between amorphous and crystalline ices, for instance. This phenomenon stems from a balance between absorption and scattering and must be taken into account in detailed modeling of spatially-resolved infrared spectroscopy of ices, including CO and CO$_2$. }
   {}


   \titlerunning{The changing shape of ice bands across protoplanetary disks}
    \authorrunning{L. Martinien et al.}
   \maketitle
   
%

\section{Introduction}

Ices form in the cold regions of protoplanetary disks and are involved in many processes leading to planetary formation. They can facilitate the formation of complex organic molecules \citep{Boogert2015} and are the bulk reservoir of volatile species \citep{Pontoppidan_2014}, playing a key role in the chemical evolution from the molecular cloud to the protoplanets 
\citep{McClure_2023}. Moreover, the role of ices (H$_{2}$O, CO, CO$_{2}$) as facilitator of grain growth has long been assumed, although the situation remains unclear following recent laboratory experiments on ice sticking properties 
\citep[e.g.,][]{Musiolik_Wurm_2019}.

While the interpretation of the spectral shape of ice features can be done when observed in absorption in the interstellar medium \citep{Whittet1997}, 
through molecular clouds \citep{Dartois2024}, circumstellar envelopes \citep{Pontoppidan2005} or in single scattering off optically thin debris disks \citep{Kim2024}, the situation is more complex in optically thick protoplanetary disks, devoid of dusty envelopes, where both multiple scattering and absorption contribute to the emerging spectrum. A full radiative transfer model is then required to disentangle the respective contributions of scattering albedo and overall opacity \citep{Sturm2023_HH48_ices, Sturm2023_HH48_modeling}. Recent studies demonstrated the presence of a significant shift, both in terms of wavelength and shape, in the spectra of H$_{2}$O, CO, and CO$_{2}$ ice bands resulting from a balance between absorption and scattering \citep{Sturm2023_HH48_ices, Dartois_2022, Martinien_2024}. Beyond confirming the presence of icy mantles
onto solid particles, near-infrared spectroscopic observations can provide critical constraints on the size distribution of dust grains \citep[e.g.][]{Dartois_2022} as well as the role of scattering over extinction, a fact often neglected so far. This is a key piece of information missing in the general context of grain growth. 

The $\sim$\,3\,\mic\, spectral signature of water ice is commonly detected in the clouds of star-forming regions \citep{Boogert2015}. Although ices are also predicted to be ubiquitous in protoplanetary disks and are often inferred from sub-mm molecular line mapping tracing sublimation \citep{Miotello2023}, they have been directly detected in only a handful of cases to date. On the one hand, this has been done through imaging and photometry in low-inclination disks, where the ice band is observed purely as a scattering (albedo) feature \citep{Honda_2009, Honda_2016, Betti2022}. The lack of spectroscopic information in this case precludes a detailed analysis of the ice properties. On the other hand, water-ice bands have been detected spectroscopically in a few highly inclined disks where the glare of the central star is naturally blocked by the disk itself: HK Tau B and HV Tau C \citep{Terada_2007}, PDS 453 \citep{Terada_2017} and HH 48 NE \citep{Sturm2023_HH48_ices}. These highly inclined disks are suitable candidates to study the spatial distribution of ices. However, these observations were spatially unresolved so that the spatial distribution of ices in disks is currently not well constrained. Furthermore, these spectra reveal deep and saturated features dominated by multiple scattering, precluding a robust estimation of the amount of ices. Arguably, the ideal systems to study the spatial distribution and properties of ices are so-called grazing-angle disks (e.g., MY Lup \citep{Avenhaus_2018}, PDS 111 \citep{Derkink_2024}, PDS 453 \citep{Martinien_2024}). In this specific geometry, a central point source is well-defined and the inclination of disks is sufficiently high so that the upper layers of the disk lay in our line-of-sight to the central star, but not so high that all photons have scattered multiple times before reaching the observer. By comparing the spectrum of the central point source to those of the fainter outer regions of disks where scattered light off the disk surface dominates, it is possible in principle to solve for the absorption-vs-scattering ambiguity by taking advantage of the different scattering geometry.

Observations with the James Webb Space Telescope (JWST) yield new highlights about the study of ices through imaging spectrophotometry by obtaining spectra at each spaxel instead of an integrated spectrum over the whole disk. 
These new observations will make it possible to study water-ice spectra at different locations in disks. In this paper, we present a first investigation of the influence of the viewing geometry and spatial location on the spectral shape of the water-ice band.


\section{Modeling setup}
\label{sec:modeling}

We constructed a fiducial disk model, without the goal of matching any disk in particular, to explore the importance of the viewing angle and light propagation path in the shape of the H$_{2}$O, CO, and CO$_{2}$ ice spectra, the most common ice species observed in protoplanetary disks. In this section, we describe the set of parameters used to model the different spectra. Our model is based on the one described in \citet{Woitke2016}  in which they propose a combination of standard parameters in order to model protoplanetary disks around T Tauri stars in the context of the DIANA project. 

We used the MCFOST radiative transfer code \citep{Pinte_2006} to produce near-infrared images at different wavelengths taking into account full anisotropic multiple scattering. We assumed a T Tauri star, located at 140\,pc, with an effective temperature of 4000\,K and a stellar luminosity of 1\,$L_\odot$, corresponding to a mid-K-type star. We approximate the stellar spectrum as a blackbody to ensure that all spectral features in our model are due to the disk itself.

We considered a single-zone axisymmetric disk with a spatially uniform dust population that follows a surface density profile extending smoothly to the outer radius and in hydrostatic equilibrium. We consider inclinations ranging from 60\degr\ to 90\degr . The disk structure follows the set of parameters in Table 3 of \citet{Woitke2016}. Since the disk is highly optically thick far into the infrared, we assumed that the dust grains are in radiative equilibrium and local thermodynamic equilibrium.

However, our model differs slightly from the one of \citet{Woitke2016} in its assumptions about the composition of dust grains. Instead of having a mixture of Mg$_{0.7}$Fe$_{0.3}$SiO$_{3}$ and amorphous carbon, we used a mixture of two other solid species: 30\% of amorphous water ice, as described in \citet{Li_1998}, and 70\% of a mixture of silicates and graphite developed by \citet{Draine_1984} to match the interstellar dust properties including a porosity of 25$\%$. All proportions are in volume fraction. The proportion of water ice was chosen so that the band is deep enough to clearly see the effects \citep[see, e.g.,][]{Martinien_2024}. We used the Bruggeman rule effective medium theory to compute the effective refractive index with the Distribution of Hollow Spheres \citep{Min_2005}. The grain size distribution is defined between $a_{min}=0.05\,\mu$m and $a_{max}=3000\,\mu$m with a $n(a) \mathrm{d}a \propto a^{-3.5} \mathrm{d}a$ distribution.

We removed all water ice inside 1.4 au to mimic water ice sublimation since the midplane temperature is higher than 90 K. 
We assumed dust is well mixed in the disk without settling. \rev{These two assumptions do not represent the complete physical picture, where ice would be removed from parts of the upper layers of the disk where temperatures are high enough and because of photodesorption. Moreover, vertical settling is expected to remove large grains from the upper layers, also impacting the scattering process. However, we show that spectra do present significant differences even for uniform dust mixtures. Therefore, our modeling strategy underestimates the spectral diversity observed in the ice bands in a disk compared to a more realistic model, i.e., which includes vertical settling and ice removal in the upper layers. 
This will be explored in a future study}.

\section{Results}

We extracted the spectra of the disk from synthetic images in total intensity between 2.0 and 4.0\,$\mu$m at two different locations: at the central point source and at the outer edge of the disk. We applied a convolution by a wavelength-scaled Gaussian kernel as a simple approximation of the JWST point spread function but the results are generic and PSF-independent. For both extractions, we defined an extraction zone of 0\farcs2$\times$0\farcs2. The extraction zone of the central point source is centered around the brightest pixel of the image, whereas the extraction zone at the edge of the disk is centered around the brightest pixel of the upper nebula at a projected distance from the center of $\approx$0\farcs5, or 70\,au at 140\,pc.
The spectra were normalized by a third-order polynomial fit to the continuum, using the 2.0-2.3\,$\mu$m and 3.7-4.0\,$\mu$m regions to avoid the water ice band and set the continuum baseline. The relative depth as well as the position of the minimum of the water-ice bands were then obtained by a third-order polynomial fit of the continuum normalized spectra. We refrain from converting this normalized depth into an optical depth, as is often done in ice studies, because the physical meaning of the latter is only applicable to a situation where an icy foreground material absorbs light from a distant background source. Instead, in protoplanetary disks, photons have multiple paths to the observer, making the notion of optical depth ambiguous at best.

    \begin{figure}[htb]
        \centering
        \includegraphics[width=0.45\textwidth]{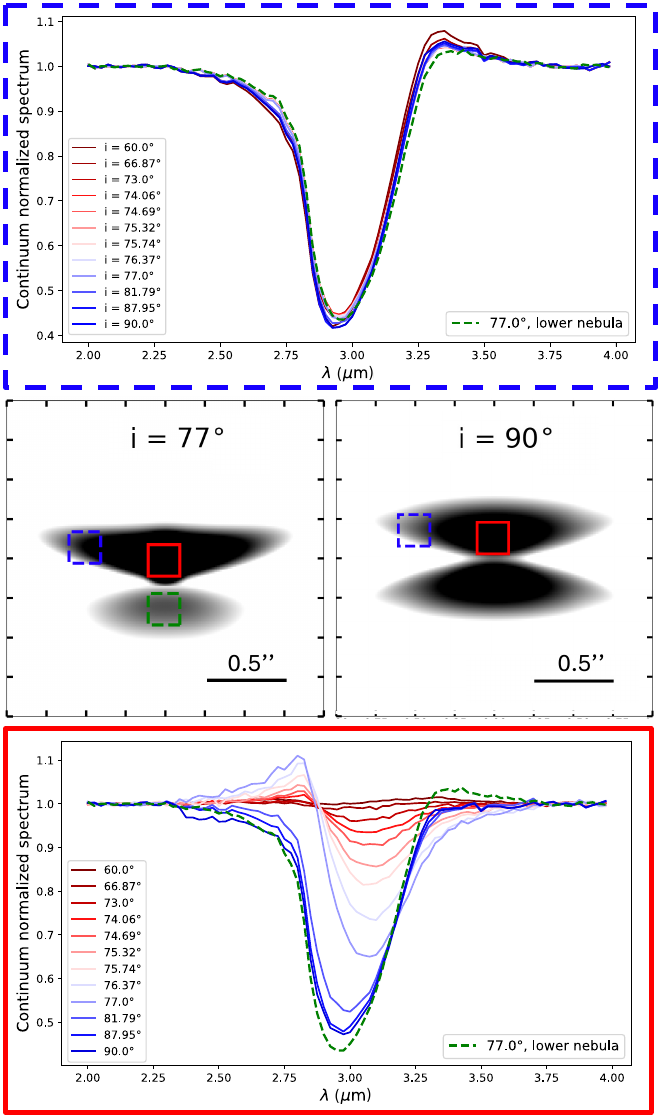}
        \caption{Synthetic spectra (top and bottom panels) and scattered-light images (middle panels) of our generic model at two different inclinations (grazing angle and edge-on). Images at 3.0\,\mic\,
        are shown with a logarithmic color scale. The squares correspond to the extraction zones in the edge of the disk (dashed blue line) and in the central point source (solid red line) with the associated spectra at different inclinations (between 60\degr and 90\degr). A spectrum of the lower nebula in the grazing angle geometry is shown in dashed green line in the top and bottom panels.}
        \label{fig:IMAGES}
    \end{figure}

    \begin{figure}[htb]
        \centering
        \includegraphics[width=0.45\textwidth]{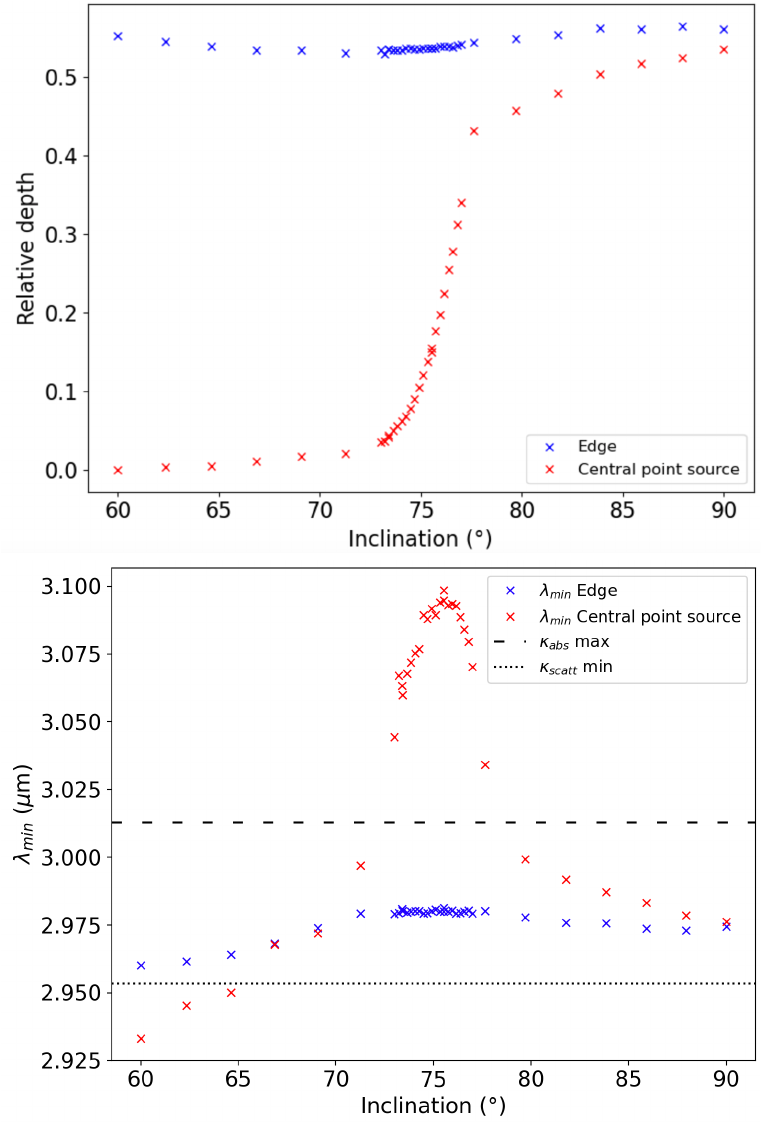}
        \caption{Relative depth and wavelength of band minimum (top and bottom panels, respectively) of the water-ice bands as a function of inclination. The dashed and dotted black lines in the bottom panel represent the wavelengths at which $\kappa_{abs}$ and $\kappa_{scatt}$ reach their minimum and maximum value across the water-ice band, respectively (see Fig.\,\ref{fig:albedo_opacity}).}
        \label{fig:depth_min_lambda}
    \end{figure}

In Fig. \ref{fig:IMAGES}, we present the spectra obtained at the central point source and at the edge of the disk. It is clear that the spectral shape as well as its dependence on inclination vary between the two positions. Below, we report the differences in terms of band shape and wavelength of the band minimum.

The shape of the water-ice bands differs according to position in the disk. This striking difference offers a strong diagnostic tool. At the disk edge, the water-ice band is asymmetric with a \rev{weak} red bump that rises above the continuum. This is never the case at the position of the central point source. \rev{This small bump might not be physical but instead due to our continuum fitting method. Indeed, by changing the wavelength range where the continuum is estimated, this bump can progressively disappear. However, we decided to set the baseline far from the band so as to not overlap the edges of the band, and in the same way for each inclination to ensure consistency.}
At the point source, the water-ice bands are symmetric at low inclinations (between 60\degr\ and 73\degr) and high inclinations (between 82\degr\ and 90\degr). However, at so-called  grazing-angle inclinations (between 74\degr\ and 77\degr\ for this model), the water-ice bands is also asymmetric but with a pronounced blue bump that rises above the continuum. \rev{In this case, the blue bump is undoubtedly physical and is clearly observed in pure spectra without normalization.}
We emphasize that the notion of grazing-angle depends on the disk geometry \rev{(e.g. scale height, flaring index) and mass} and thus does not imply a unique range of inclinations. The results presented here are generic, but the correspondence between the exact shape of the band and a given inclination is valid only for the disk model used here. We also extracted the same spectra for a very low-inclination disk (30\degr) in which water-ice bands follow the same pattern as those at our lowest inclinations, confirming that we have explored the full diversity in our model. 

The top panel of Fig. \ref{fig:depth_min_lambda} shows the relative depth to the continuum for both spectra at each inclinations. The range of relative depth to the continuum is also wider for the central point source. At the edge of the disk, the relative depths are marginally affected by the inclination and are in the same range of values ($\sim$55$\%$ to $\sim$56$\%$) whereas the values for the central point source are more widely spaced ($\sim$0$\%$ to $\sim$53$\%$). Indeed, with decreasing inclination angles, the line-of-sight to the star and the central (ice-free) disk region becomes increasingly optically thin and, consequently, the ice feature progressively disappears.

Another difference between spectra at the edge of the disk and at the central point source is the wavelength at which the minimum of the water-ice band is located. The bottom panel of Fig. \ref{fig:depth_min_lambda} shows the wavelength of the band minimum ($\lambda_{min}$) for each inclination. As for the relative depth to the continuum, the range of $\lambda_{min}$ throughout different inclinations is broader at the central point source than at the edges. $\lambda_{min}$ shifts to longer wavelengths from the lowest inclinations to the grazing-angle inclinations, from $\lambda_{min}\sim$2.93\,\mic\, to $\sim$3.10\,\mic\, at an inclination of $\sim$75\degr. Then, $\lambda_{min}$ gradually decreases back to $\sim$2.98\,mic\, for the highest inclinations.
For the spectra at the edge of the disk, this minimum position shifts only slightly between $\sim$2.96\,\mic\, for the lowest inclinations and $\sim$2.97\,\mic\, for the highest inclinations.

We also extracted spectra at the edge of the lower nebula as well as below the central point source, i.e, on the bottom surface of the disk. These water-ice bands present the same behavior and trends as those extracted at the edge of the upper nebula and at the central point source respectively with the notable exception of the grazing angle geometry. In that case, the spectrum of the lower nebula is markedly different from that of the central point source and is instead very similar to the edge of the upper nebula (dashed green line in Fig. \ref{fig:IMAGES}).  
We also calculated the integrated spectra over the whole image. In this case, the spectra are dominated by the central point source with a slight additional component at shorter wavelengths. In the next section, we interpret the various results that our models have yielded.

\section{Discussion and conclusions}

\begin{figure}[htb]
        \centering
        \includegraphics[width=0.45\textwidth]{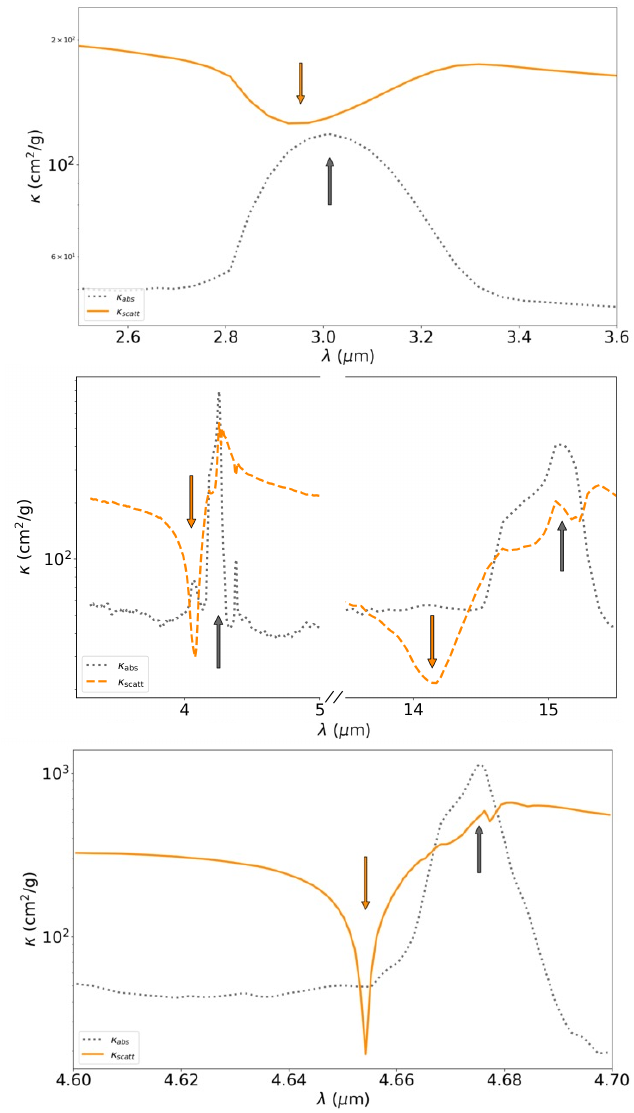}
        \caption{Scattering and absorption opacities of the dust populations in selected wavelengths regions. The top panel corresponds to the dust population used in the model presented in Section 2, whereas the middle and bottom panels correspond to pure CO$_2$ \citep{Gerakines_2020} and CO ices \citep{Palumbo_2006}, respectively.}
        \label{fig:albedo_opacity}
    \end{figure}

The observed behavior of water-ice bands can be explained by the wavelength dependency of the scattering and absorption opacities of the mixture used in our model, see Fig. \ref{fig:albedo_opacity}~(top panel). To disentangle which process dominates, we measure the shift in wavelength of the absorption band minimum. This shift is due to a difference in extremum position in the opacity curves. For the dust model we adopted, the maximum of the absorption curve is centered around 3.0\,\mic\, whereas the minimum of the scattering curve is centered around 2.9\,\mic\,. We divide inclinations into 3 ranges: low inclinations (60\degr - 72\degr), grazing-angle inclinations (73\degr - 77\degr) and high inclinations (78\degr - 90\degr). For our fiducial model, the  inclination range of grazing angles correspond to a regime of rapid increase in the depth, and shift in the minimum wavelength, of the water-ice band for the central point source. Again, this is a generic result, but the exact range of grazing-angle inclinations is model dependent. 
   
The two bumps observed on the red and blue sides of spectra extracted at the edge and central point source respectively are due to enhanced scattering. Past studies have suggested that components on either side of the main water-ice band are an indicator of grain growth \citep{Leger1983, Boogert2015, Dartois2024}. To test this, we considered an alternative model that only contains small grains between 0.03 and 0.25\,\mic\, (Fig. \ref{fig:small_grains}). 
\rev{In this case, the red bumps are still present, and even more pronounced, at the two locations, although we cannot exclude that they are artefacts of the continuum normalization.} 
If real, this is caused by the scattering opacity and single scattering albedo profiles that both have a peak on the red side of their respective minimum \citep{Inoue_2008}, see bottom panel of Fig. \ref{fig:small_grains}. On the other hand, \rev{at the central point source,} the blue bump is now absent, confirming that it is caused by scattering off large grains. \rev{However, a small blue bump appears at the edge where spectra are scattering-dominated, although again it could be due to our continuum fitting method.}

The behavior of the minimum wavelength can be decomposed according to the location in the disk: edge versus central point source. At the edge of the disk, the minimum position of the spectra for all three ranges of inclinations lie in a narrow range around $\sim$2.96\,\mic.
In other words, scattering dominates at all inclinations. However, at the central point source, the minimum position shifts across the three ranges of inclinations. At low inclinations, this minimum is between $\sim$2.93\,\mic\ and $\sim$2.99 \,\mic\, where scattering dominates. At grazing-angle inclinations, the minimum shifts all the way to 3.1\,\mic\, before shifting back towards $\sim$2.975\,\mic\, at high inclinations. This means that grazing-angle inclinations correspond to a regime in which absorption dominates over scattering whereas both high and low inclinations correspond to a regime where scattering dominates, \rev{except for the lowest inclinations where the direct starlight reaches the observer resulting in no ice bands.} 

As mentioned in Sect. \ref{sec:modeling}, we used amorphous water-ice in our model. However, several studies of ices have suggested that the minimum wavelength depends on the type of ice considered. 
Both laboratory data and astronomical observations show that a peak position around $\sim$2.9\,\mic\ is due to amorphous water ice while a peak position around $\sim$3.1\,\mic\ is due to crystalline water ice \citep{Smith_1988, Boogert_2008, Terada_2012, Boogert2015}.  
Furthermore, \citet{Hudgins_1993} and \citet{Mastrapa_2009} showed that the water-ice band depends on temperature. 
Here we show that the position of this minimum does not solely depend on the type of ice but also on the inclination of the system as well as where the spectra are extracted from in disks. Indeed, the amplitude of the wavelength shift we observed on the central point source is comparable to the difference between the amorphous and crystalline ices.

The relative depth to the continuum is also more variable for the central point source spectra than for the edge spectra. In both cases, spectra reach a saturation for the highest inclinations making it impossible to determine the amount of ice in disks that are too highly inclined. This phenomenon was reported both in models \citep[e.g.,][]{Pontoppidan2005, Sturm2023_HH48_ices, Martinien_2024} as well as in observations
of edge-on disks by \citet{Aikawa2012}. 

We interpret the different behaviors between the two locations as being related to the way photons propagate through the disk. In terms of optical depth, at the lowest inclinations, the stellar photosphere is barely extincted by the disk and remains fully visible. The observer's line of sight to the central star is essentially dust-free and photons have a negligible absorption component. However, some light scatters off the disk surface at close projected separations, resulting in a scattering-dominated water-ice band on the central source. With increasing inclination towards grazing-angle geometry, the line-of-sight optical depth also increases and the disk progressively masks the star. The stellar light then has to propagate through the upper surface of the disk leading to significant absorption and a non-negligible scattering component. Grazing-angle inclinations is the only geometry for which the optical depth is such that the light transmitted to the observer is both absorbed and scattered. When the inclination increases further, the line of sight to the star is optically thick. Stellar photons cannot reach the observer directly and most them are absorbed. The absorption component then becomes negligible compared to scattering as photons must undergo multiple scattering events before reaching the observer, hence leading to a spectrum dominated by scattering and a rapidly saturated water-ice band.

On the other hand, at the edge of the disk and in the lower nebula, photons must scatter multiple times before reaching the observer, leading to deeper bands at any inclinations. 
\rev{Indeed, photons are scattered following the same path from the central star towards the edge of the disk. The slight differences in terms of relative depth at the edge observed in the top panel of Fig. \ref{fig:depth_min_lambda} result from the path between the edge and the observer which depends on the optical depth of the disk and therefore, the inclination.}

The phenomena we observe in our model can be generalized to other types of ices, so long as there is a shift between the real part and the imaginary part of the refractive indices. This is observed for all types of water ice, whether 
amorphous or crystalline \citep{Hudgins_1993, Warren_2008}. 
This is also the case for CO$_{2}$ and CO ice bands. 
The middle and bottom panels of Fig. \ref{fig:albedo_opacity} show opacity curves for pure CO$_{2}$ and CO ice, respectively. Similarly to water ice, shifts between the extrema of the scattering and absorption opacity curves are found for the two CO$_{2}$ bands at 4.2\,\mic\ and 15.2\,\mic, as well as for the CO ice band at 4.6\,\mic. We therefore predict similar shifts in wavelengths and band shapes for these species as a function of inclination and spatial location in disks. 
We note that the presence of a blue bump in grazing angle systems has been independently predicted in radiative transfer models \citep[for CO$_2$, see Fig.\,14 in][]{Dartois_2022}. Recent observations of CO and CO$_{2}$ ices in a grazing-angle disk in Orion, as suggested by optical imaging of the system \citep{Smith_2005}, further confirm our predictions \citep{Potapov_2025}.

Water-ice spectra observed in protoplanetary disks depend on the inclination of the system as well as the location in the disk from which they are extracted. We emphasize again that our model has the exact same dust properties (composition, ice fraction) at all locations \rev{except in the inner 1.4\,au where the ice is removed}, yet we have demonstrated that depending on inclination and location, significantly different water-ice band morphologies are expected. Both dependencies can be explained by a balance between absorption and scattering leading to differences in shape, symmetries, relative depth, and minimum position. These results have an impact on  models and on the interpretation of observations. Models must account for scattering to avoid misinterpretation of ice band behaviors. Similarly, observations of ices in protoplanetary disks must carefully consider these effects to accurately constrain ice properties within the disk. On the other hand, the very distinct behavior between center and edge of disks, as well as the sensitivity and spectral resolution of JWST's integral field units, offer a new and powerful opportunity to further understand the properties of ices in protoplanetary disks. This study will also be relevant in the context of the SPHEREx mission which will produce 1-5\,\mic\, spectra of a large number of disk-bearing sources in most star-forming regions \citep{Melnick_2025}.

\begin{acknowledgements}
This project has received funding from the European Research Council (ERC) under the European Union’s Horizon Europe research and innovation program (grant agreement No. 101053020, project Dust2Planets, PI: F. M\'enard).
\end{acknowledgements}

\bibliography{BibLau.bib}

\begin{appendix}

\section{Spectra for a model containing only small grains}

\begin{center}
    \begin{figure}[h]
        \centering
        \includegraphics[width=0.45\textwidth]{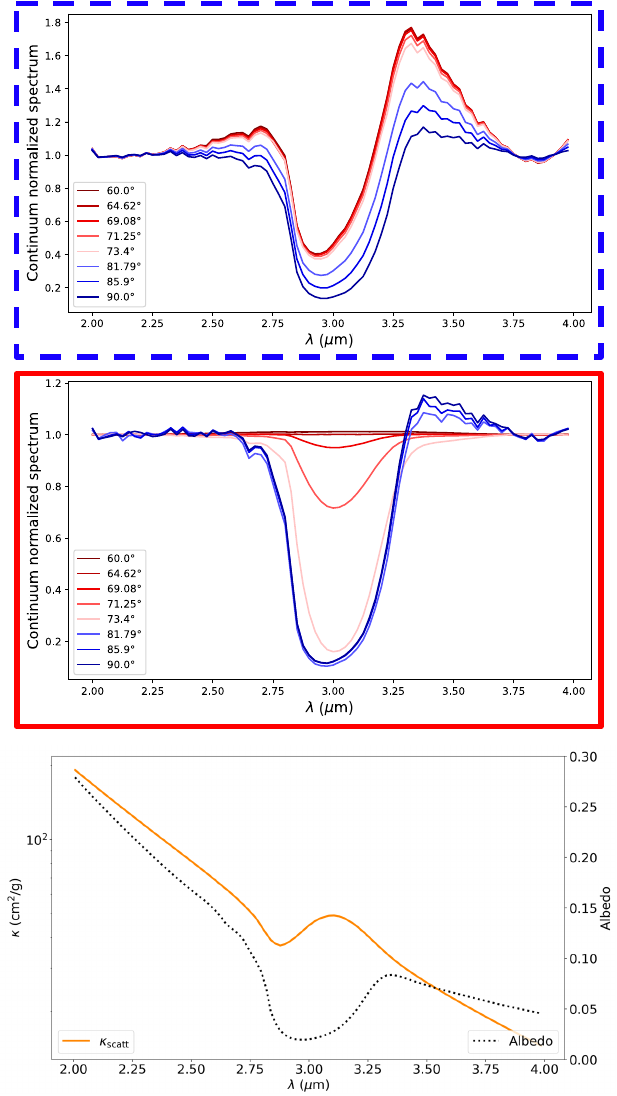}
        \caption{Synthetic spectra (top and middle panels) as well as opacities and albedo curves (bottom panel) of an alternative model at different inclinations (between 60\degr and 90\degr). The top panel corresponds to the edge of the disk whereas the middle panel corresponds to the central point source.}
        \label{fig:small_grains}
    \end{figure}
\end{center}

Figure \ref{fig:small_grains} presents spectra at different inclinations for an alternative model that only contains small grains between 0.03 and 0.25 \mic\,. For consistency, the zones have been extracted and the spectra normalized following the same procedure as for our generic model.

\end{appendix}


\label{LastPage}
\end{document}